\documentclass{pasj00}
\draft

\begin{document}
\SetRunningHead{HAYAKAWA ET AL.}{XMM-NEWTON OBSERVATION OF A~1060}
\Received{2000/12/31}%{yyyy/mm/dd}
\Accepted{2001/01/01}%{yyyy/mm/dd}

\title{Detailed XMM-Newton Observation of the Cluster of Galaxies Abell 1060}

%%% begin:list of author
\author{
Akira \textsc{hayakawa}\altaffilmark{1},
Akio \textsc{hoshino}\altaffilmark{1},
Manabu \textsc{ishida}\altaffilmark{1}, 
Tae \textsc{furusho}\altaffilmark{2},
Noriko \textsc{y. yamasaki}\altaffilmark{2}, \\
and \\
Takaya \textsc{ohashi}\altaffilmark{1}}
\altaffiltext{1}{Department of Physics, Tokyo Metropolitan University, \\ 
1-1 Minami-Osawa, Hachioji, Tokyo 192-0397}
\altaffiltext{2}{Institute of Space and Astronautical Science, 
Japan Aerospace Exploration Agency, \\
3-1-1 Yoshinodai, Sagamihara, Kanagawa 229-8510}
\email{E-mail(MI): ishida@phys.metro-uac.jp}
%%% end:list of authors

%% `\KeyWords{}' always has to be placed before `\maketitle'.
\KeyWords{galaxies: clusters: individual (Abell 1060) -- galaxies:
intergalactic medium -- X-ray: galaxies} %Do NOT move this preamble from
%here!
\maketitle

\begin{abstract}
We present results from the {\it XMM-Newton} observation
of the non-cooling flow cluster A~1060.
Large effective area of {\it XMM-Newton} enables us to investigate
the nature of this cluster in unprecedented detail.
From the observed surface brightness distribution, 
we have found that the gravitational mass distribution
is well described by the NFW profile 
but with a central density slope of $\sim$1.5.
We have undoubtedly detected a radial temperature decrease
of as large as $\sim$30\% from the center
to the outer region ($r \sim13'$),
which seems much larger than that expected 
from the temperature profile averaged over nearby clusters.
We have established that the temperature
of the region $\sim 7'$ southeast of the center 
is higher than the azimuthally averaged temperature of the same radius
by $\sim$20\%.
Since the pressure of this region already reaches equilibrium
with the environment,
the temperature structure can be interpreted as having been produced
between $4 \times 10^{7}$~yr (the sound-crossing time) 
and $3 \times 10^{8}$~yr (the thermal conduction time) ago.
We have found that the high-metallicity blob located 
at $\sim 1.\!'5$ northeast of NGC~3311 is more extended
and its iron mass of $1.9 \times 10^{7}$~M$_{\odot}$
is larger by an order of magnitude
than estimated from our {\it Chandra} observation.
The amount of iron can still be considered as being injected 
solely from the elliptical galaxy NGC~3311. 

\end{abstract}

%\newpage

%%%%%%%%%%%%%%%%%%%%%%
\section{Introduction}
%%%%%%%%%%%%%%%%%%%%%%

Clusters of galaxies, being the largest virialized systems in the
universe, are filled with the intracluster medium(ICM), which consists
of X-ray emitting hot plasma with a typical temperature of a few
$\times 10^{7}$~K.  The ICM gives us rich information about the
structure and evolution of clusters of galaxies.  X-ray Spectra of ICM
immediately reveal its temperature and metal abundance, and the ICM
surface brightness shows us ICM density distribution.  Through the
assumption of hydrostatic equilibrium, furthermore, it is possible to
measure the distribution of the gravitational mass in clusters.

The cluster mass distribution is an important observational clue in
determining the nature of the dark matter and in constraining the
structure formation scenario. In particular, the CDM model predicts a
significant central cusp in the dark matter profile

\citep{1996ApJ...462..563N},
in contrast to the flat core profile described by the $\beta$
model. High-sensitivity X-ray study is a powerful method to look into
the gravitational mass distribution (e.g. \cite{2003ApJ...590..225C}). 
For clusters characterized by cool central cores,
presence of the bright central galaxy greatly hampers the mass
determination for the pure cluster component. Therefore, detailed
study of a non cool-core cluster has a significant importance.

Recent observational studies of the clusters with {\it Chandra} and
{\it XMM-Newton}, with their powerful imaging capability and large
effective area, unveiled a lot of new interesting features.
\citet{2003ApJ...596..190M} found an unusual X-ray morphology, blob
and hole like features, in the central dense region of the 2A~0335+096
cluster.  \citet{2000ApJ...541..542M} detected sharp brightness edges,
called cold fronts, with lower temperature inside the subcluster in
A~2142 cluster.  \citet{2003ApJ...596..181F} found two
high-metallicity blobs located symmetrically with respect to the
center of the poor cluster AWM~7.  \citet{2005A&A...433..101P}
investigated temperature profiles of 13 nearby cooling flow clusters
observed with {\it XMM-Newton}, and found that the temperature
decreases by $\sim 30$\% between 0.1~$r_{vir}$ and 0.5~$r_{vir}$.
This feature is in good agreement with the result of
\citet{1998ApJ...503...77M} who found that the temperature of 30
nearby clusters show a significant decline at large radii.  Recent
{\it Chandra} observation of A~1060 revealed that the central two
elliptical galaxies emit X-ray \citep{2002ApJ...578..833Y}, and the
temperature and the abundance distribution are unexpectedly
inhomogeneous \citep{2004PASJ...56..743H}.

A~1060 is a nearby cluster ($z = 0.0114$) characterized by a smooth
and symmetric ICM distribution.  The ASCA and ROSAT observations found
that the temperature and the abundance showed very uniform
distribution \citep{1996PASJ...48..671T,2001PASJ...53..421F}.  For
these reasons, this cluster was thought to be in a highly relaxed
state, and has been used to estimate the gravitational mass profile
which can be directly compared with the results of hydrodynamic
cluster simulation \citep{2000ApJ...535..602T}.

In this paper, we report the results of {\it XMM-Newton} observation
of the A~1060 cluster.  Based on the surface brightness distribution, we
have computed the total gravitational mass of the cluster.  We
confirmed the existence of the high-metallicity blob discovered by our
{\it Chandra} observation but with a larger iron mass content.  Using
the high spatial resolution and the large effective area, we searched
for a sign of a sub-cluster merger from a two-dimensional
temperature map.  Throughout this paper, we assume
$H_{0}=75$km~s$^{-1}$~Mpc$^{-1}$ with $q_{0}=0.5$.  Accordingly, an
angular size of $1'$ corresponds to 13~kpc.  The solar number
abundance of iron relative to H is assumed to be $4.68\times 10^{-5}$
\citep{1989GeCoA..53..197A}.  We employ Galactic absorption as
$N_{\rm H}=4.9 \times 10^{20}$cm$^{-2}$.

%%%%%%%%%%%%%%%%%%%%%%%%%%%%%%%%%%%%%%%%
\section{Observation and Data reduction}
%%%%%%%%%%%%%%%%%%%%%%%%%%%%%%%%%%%%%%%%

The {\it XMM-Newton} observation of the A~1060 cluster of galaxies
was carried out on 2004 June 29 for a total exposure time of 64~ksec.
All the EPIC cameras were operated in the Full frame mode
with the medium filter inserted.
Data reduction and analysis were performed with SAS version 6.0 and CIAO 3.0.
Unfortunately, part of the data was affected by high background flares.
In order to remove them, we adopted the following data cleaning process.
We made light curves in the bands 10-12~keV for MOS and 12-14~keV for pn,
where the signals are dominated by background events,
and then we removed the time interval when the count rate 
is larger than 0.3 c~s$^{-1}$ for MOS and 0.8 c~s$^{-1}$ for pn.
Then, we selected the photon events with pattern 0-12 for MOS and 0-4 for pn, 
and flag=0
\footnote{http://xmm.vilspa.esa.es/sas/current/doc/evselect/}
.
After these selections, some 35.9, 36.4, and 25.2~ksec of good time intervals
remained for MOS1, MOS2, and pn, respectively.

\begin{figure}[htb]
  \begin{center}
%    \FigureFile(80mm,80mm){final-fig/X-ray-img.eps}
    \FigureFile(80mm,80mm){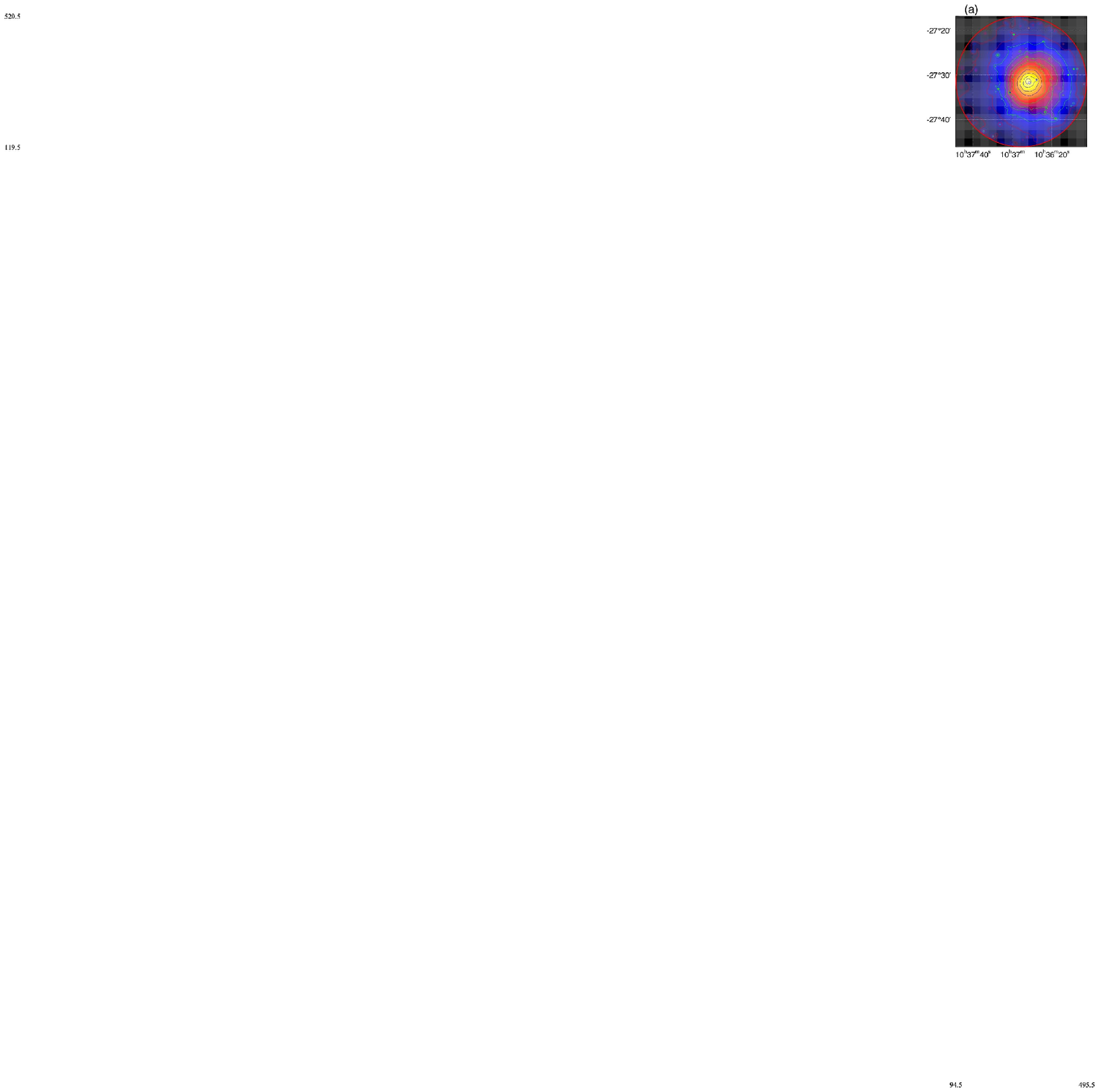}
    \FigureFile(80mm,80mm){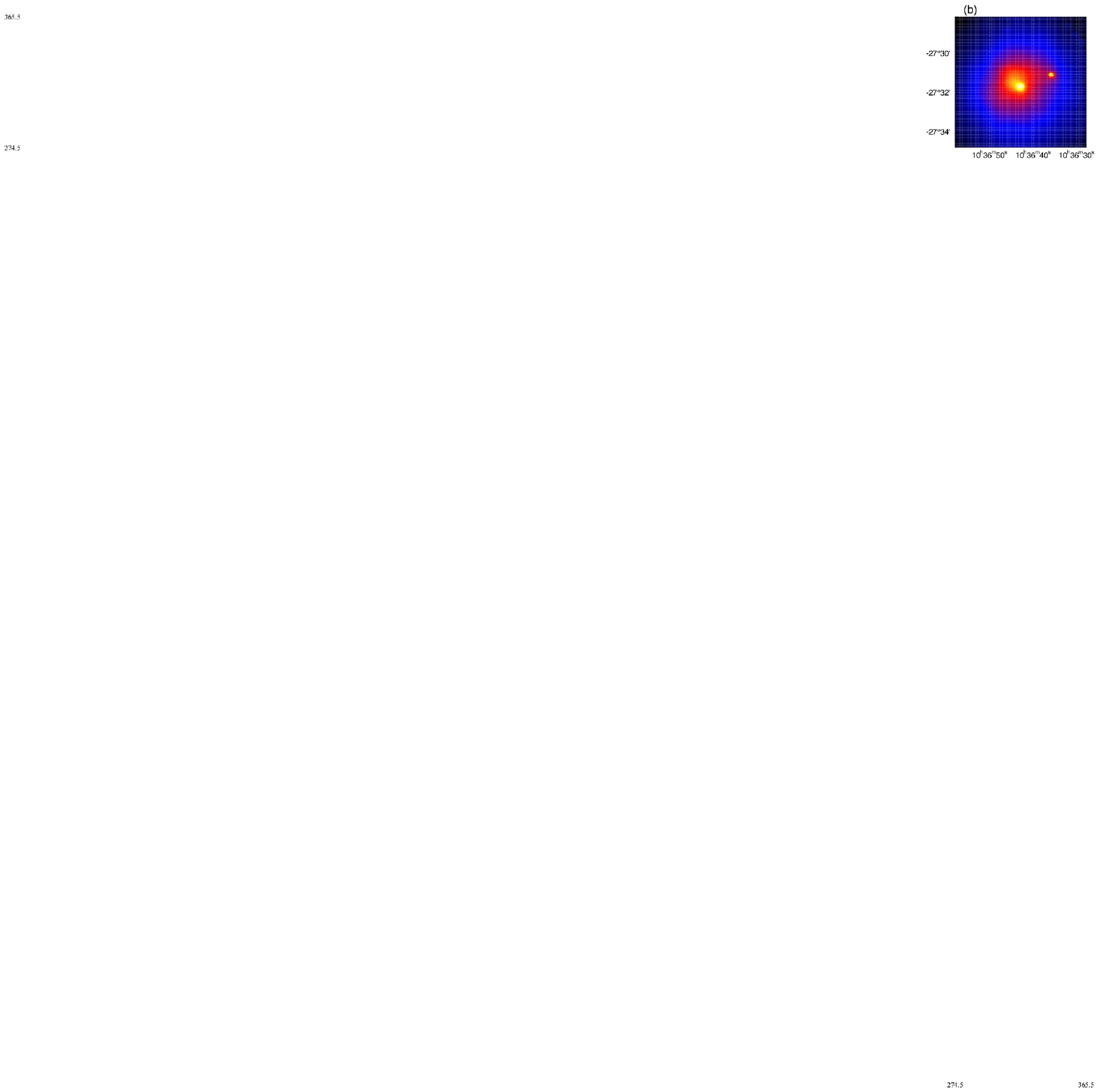}
  \end{center}
\caption{(a) Adaptively smoothed MOS1 X-ray image 
of the $30'\times 30'$ region, overlaid with the intensity contours. 
The contours were logarithmically scaled and divided into 20 spaces
 between $10^{-5}$ and  $10^{-3}$counts~s$^{-1}$ range. 
The red circle represents {\it XMM-Newton} field of view.
(b) Close up view in the central $8'\times 8'$ region.
}
\label{fig:X}
\end{figure}

The blank sky data sets of \citet{2003A&A...409..395R} were used 
to subtract the background of our data sets.
We extracted the background events using the same selection criteria as above.
The source-to-background count-rate ratio was calculated 
from the count rates in the bands 10-12 and 12-14~keV for MOS and pn 
(see \cite{2004A&A...414..767K}), which result in 1.132, 1.250, and 1.200 
for MOS1, MOS2, and pn, respectively.
After subtracting the blank sky data, being scaled by these factors, 
from the source data sets, 
we created X-ray images for each detector separately.
Figure \ref{fig:X} (a) and (b) show the X-ray image of MOS1 
in the band 0.8-8.0~keV. 
The telescope vignetting effect had been corrected using the exposure map 
which was calculated with the SAS ``eexpmap''
\footnote{http://xmm.vilspa.esa.es/sas/current/doc/eexpmap/}
 command.
The image had been smoothed by a Gaussian with a 1$\sigma$ width 
of 4.4$''$ using the CIAO csmooth command.
The two bright elliptical galaxies (NGC~3311 and NGC~3309) 
are clearly visible in the central region of the cluster
(Fig.~\ref{fig:X} (b)).
Properties of these galaxies have been reported by \citet{2002ApJ...578..833Y}.
A blob-like high brightness region was located at northeast of NGC~3311.
Except for these central features, the whole surface brightness distribution 
remains nearly spherically symmetric.
Although the surface brightness map in Fig.~\ref{fig:X}(a)
seems elongated in the north-south direction,
the lengths of the semi-major/minor axes
of the isophote at $r\simeq 5'$, for example, 
agree within a few percents.

\begin{figure}
  \begin{center}
    \FigureFile(80mm,80mm){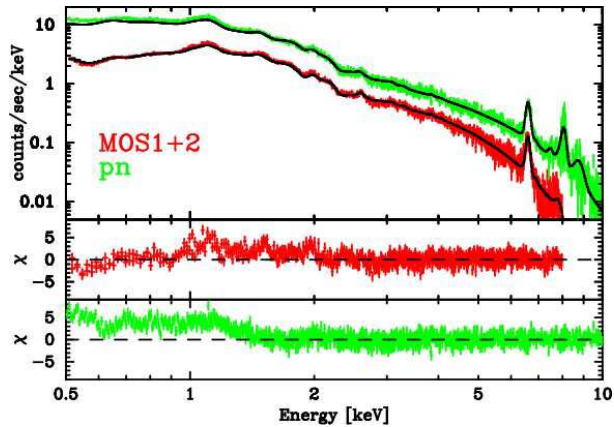}
  \end{center}
\caption{MOS1+2 and pn spectra extracted from the r$<13'$ region
fitted in the bands 0.5-8~keV and 0.5-10~keV, respectively, 
with the MEKAL model undergoing the galactic absorption.}
\label{fig:spec}
\end{figure}

We created on-source spectra of each detector within a radius of 13$'$ 
from the cluster center.
The point-like sources seen in Fig. \ref{fig:X} were removed 
using SAS ``edetect\_chain''
\footnote{http://xmm.vilspa.esa.es/sas/current/doc/edetect\_chain/}
 command.
The central two elliptical galaxies possess an extended halo
\citep{2002ApJ...578..833Y}.
Accordingly, we have excluded manually the circular regions
with radii of 25$''$ and 20$''$ centered on NGC~3311 and NGC~3309,
respectively, for accumulating spectra
in order to avoid contamination from the halo.
The blank sky spectra were subtracted from the on-source spectra
after the appropriate scalings to remove the background photons.
For pn data, there are strong fluorescence lines of Ni, Cu and Zn
at around 8~keV. 
Since these lines can not be completely removed by the background
subtraction, 
we introduce Gaussians in the spectral model
to represent these lines. %\citep{2004A&A...414..767K}.
First of all, we fitted the spectra from each detector separately
with the absorbed MEKAL model in the bands 0.8-8.0~keV for MOS1,2 and
0.8-10.0~keV for pn.
As summarized in table \ref{tab:13}, with the absorption
fixed at the galactic value of $N_{\rm H}=4.9\times 10^{20}$~cm$^{-2}$ 
\citep{1990ARA&A..28..215D}, the temperature is distributed 
in the range 3.10-3.30~keV among the three detectors.
\begin{table}
\caption{Best-fit Parameters of the MEKAL model
to the Spectra in the central $r<13'$ region}
\label{tab:13}
  \begin{center}
\begin{tabular}{lcccc} \hline 
Inst. & kT  & $Z$ & N$_{H}$ & $\chi^{2}$/d.o.f \\ 
  & [keV] & [$Z_\odot$] & [$\times 10^{20}$ cm$^{-2}$]&  \\ \hline \hline
MOS1 & $3.30^{+0.04}_{-0.04}$ & $0.41^{+0.02}_{-0.02}$ & 4.9(fix) & 1.244 \\
     & $3.23^{+0.05}_{-0.05}$ & $0.40^{+0.02}_{-0.02}$ & $6.0^{+0.6}_{-0.6}$ & 1.226 \\
MOS2 & $3.20^{+0.03}_{-0.04}$ & $0.41^{+0.02}_{-0.02}$ & 4.9(fix) & 1.366 \\
     & $3.13^{+0.05}_{-0.05}$ & $0.40^{+0.02}_{-0.02}$ & $6.1^{+0.6}_{-0.6}$ & 1.344 \\
pn   & $3.10^{+0.03}_{-0.02}$ & $0.37^{+0.01}_{-0.01}$ & 4.9(fix) & 1.475 \\
     & $3.31^{+0.02}_{-0.02}$ & $0.39^{+0.02}_{-0.02}$ & $1.2^{+0.5}_{-0.4}$ & 1.238 \\ \hline
MOS+pn & & & &  \\ \hline \hline
$>2$~keV & $3.39^{+0.04}_{-0.04}$ & $0.35^{+0.02}_{-0.02}$ 
	 & 4.9(fix) & 1.145 \\
$>0.8$~keV$^{\rm a}$ & $3.27^{+0.02}_{-0.02}$ & $0.39^{+0.02}_{-0.01}$ 
	 & 4.9(fix) & 1.250 \\ \hline
\end{tabular}
\begin{minipage}{0.6\textwidth}
{\sc Note} --- All errors are at the 90\% confidence level.\\
a: The energy bands for MOS and pn are 0.8-8~keV and 1.6-10~keV, respectively.
\end{minipage}
  \end{center}
\end{table}
If the $N_{\rm H}$ value is set free to vary, 
a similar temperature distribution ($kT$ = 3.13-3.31~keV) is obtained.
The $N_{\rm H}$ value from the pn fit, however, shifted to a lower value 
of $1.2^{+0.5}_{-0.4}\times 10^{20}$~cm$^{-2}$, 
which is significantly smaller than $4.9\times 10^{20}$~cm$^{-2}$
obtained from the radio 21~cm measurement \citep{1990ARA&A..28..215D}.
We then carried out a combined fit of the absorbed MEKAL model 
to the MOS1+2 and pn spectra
of the $r < 13'$ region in the band 2.0-10.0~keV, free from
the absorption, and simply retrieved the data points 
below 2~keV.
The result is shown in Fig.~\ref{fig:spec}.
The hydrogen column density is fixed at the galactic value, $4.9 \times
10^{20}$~cm$^{-2}.$
The best-fit parameters are summarized in table \ref{tab:13}.
The MOS and pn results are inconsistent below $\sim$1~keV, 
and the pn fit requires a strong soft excess.
A similar phenomenon has been reported from the {\it XMM-Newton} observation 
of A~1413 \citep{2002A&A...394..375P}.
According to these authors, this effect could be 
due to a true soft excess emission (e.g. \cite{2002A&A...390..397D})
and/or an artifact due to remaining calibration uncertainties.
In the case of A1060, however, such a soft excess is incompatible
with previous observations \citep{2000ApJ...535..602T}.
Accordingly, we neglect the energy band below 1.6~keV for the pn, 
and hereafter adopt the 0.8-8.0~keV and 1.6-10.0~keV bands
for the MOS and the pn, respectively.
The results of the combined fit in these energy bands are also summarized 
in table \ref{tab:13}, 
showing that the temperature of the plasma within the projected central 
$r<13'$ region is 3.27$\pm$0.02 keV with the metal abundance 
of 0.39$^{+0.02}_{-0.01}Z_\odot$.

%%%%%%%%%%%%%%%%%%%%%%%%%%%%%%%%%%%%%%%%
\section{Temperature and Abundance Distribution}
%%%%%%%%%%%%%%%%%%%%%%%%%%%%%%%%%%%%%%%%

\subsection{Radial Profile}

\begin{figure}
  \begin{center}
    \FigureFile(80mm,80mm){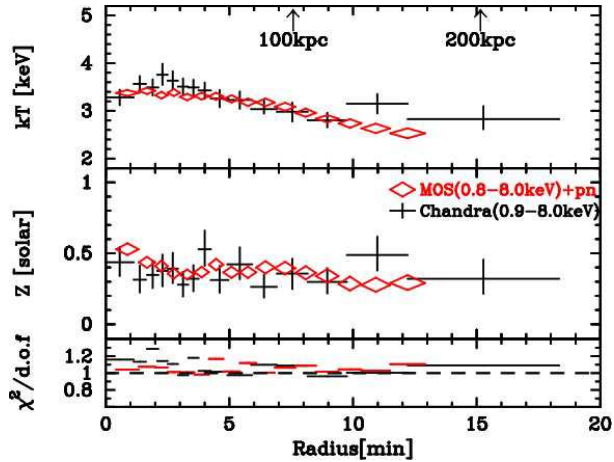}
  \end{center}
\caption{Radial distributions of the temperature (top), 
metal abundance (middle), and reduced $\chi^{2}$ value  (bottom)
of the spectral fits to the concentric annular regions.
The red diamonds show the results from the MOS spectra 
in the bands 0.8-8.0~keV (pn energy band is fixed in 0.8-8.0~keV).
The black dots show the result of {\it Chandra} observation.
The error bars are at the single-parameter 90\% confidence level.}
\label{fig:radi}
\end{figure}

We have produced radial temperature and abundance profiles 
in annuli centered on the position of NGC~3311.
The central two elliptical galaxies (NGC~3311, NGC~3309) 
and the other point-like sources are removed (\S~2).
The annuli are segmented from inner to outer radii 
so that they include at least 20,000 photons for all the detectors. 
The resulting widths of the annuli are $\sim$30$''$ or larger,
which is larger than the HPD of the XRT ($\sim$17$''$).
The MOS and pn spectra
in the bands 0.8-8.0~keV and 1.6-10.0~keV, respectively, 
are fitted simultaneously with an absorbed MEKAL model.
The absorber's hydrogen column density is fixed 
at the Galactic value of $4.9 \times 10^{20}$~cm$^{-2}$.
Figure \ref{fig:radi} shows the result 
of radial temperature and abundance distributions.
For comparison, we have overlaid the {\it Chandra} results
\citep{2004PASJ...56..743H}.
Owing to large effective area of {\it XMM},
the gradual temperature decline
from $kT =$ 3.5~keV to 2.5~keV toward the outer radii,
which is not very clear in the {\it Chandra} result,
is unambiguously detected.
Because the field of view of the {\it Chandra} observation 
was offset toward the southwest direction, the outer two annuli were 
of poor statistics and probably affected by background uncertainty.
The radial abundance increase of $\sim$20~\% in the inner 3$'$
region also becomes very clear for the first time.

We have so far adopted the position of NGC~3311
as the cluster center simply because it locates
at the apparent peak of the surface brightness,
and one may suspect that the radial profiles
shown in Fig.~\ref{fig:radi} could be significantly different
if we draw them centered on a different position.
Accordingly, as an alternative,
we have tried to define a cluster center
with the geometry of the isophotes.
We have fitted a circle to every isophote shown in Fig.~\ref{fig:X}(a)
with its central position and radius being set free to vary, and
have found that the center of the circle
fitted to the second largest isophote
($r \sim 11'$) shows the largest shift of 22$''$
from NGC~3311 in the western direction.
We have made the radial profiles again with this new center.
The resultant difference from the previous results is, 
however, quite small, which is only 5\% and 15\% at most
for the temperature and the abundance
over the entire radius range, respectively.
In particular, the 5\% random difference in the temperature
is much smaller than the systematic temperature drop from 3.5~keV
at the center to 2.5~keV at $r = 13'$.
Accordingly, we adopt the position of the galaxy NGC~3311
as the cluster center throughout this paper.

%%%%%%%%%%%%%%%%%%%%%%%%%%%%%%%%%%%%%%%%
\subsection{Two-Dimensional Distribution}
%%%%%%%%%%%%%%%%%%%%%%%%%%%%%%%%%%%%%%%%
\subsubsection{Temperature distribution}
%%%%%%%%%%%%%%%%%%%%%%%%%%%%%%%%%%%%%%%%

\begin{figure}
  \begin{center}
    \FigureFile(80mm,80mm){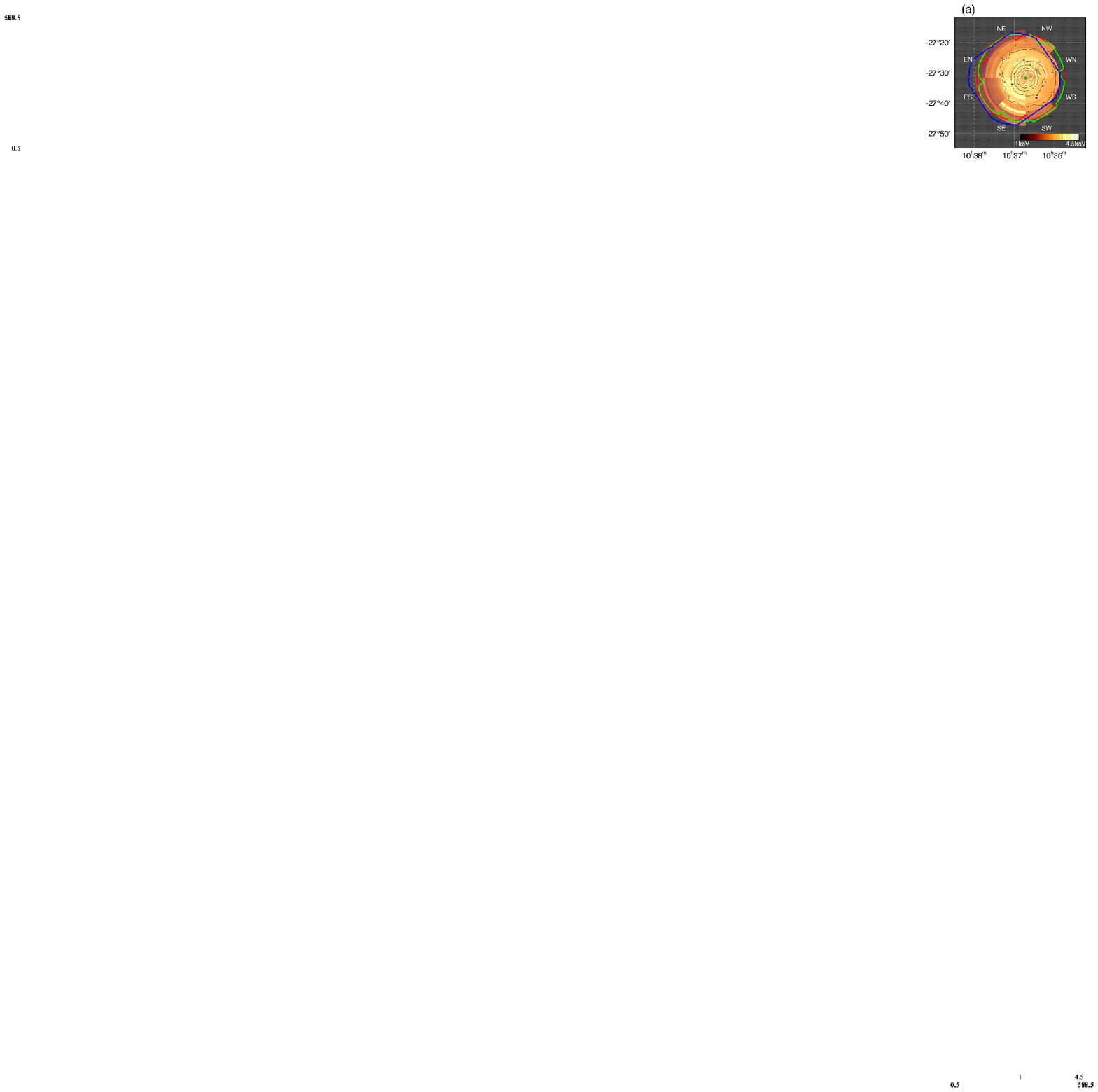}
    \FigureFile(80mm,80mm){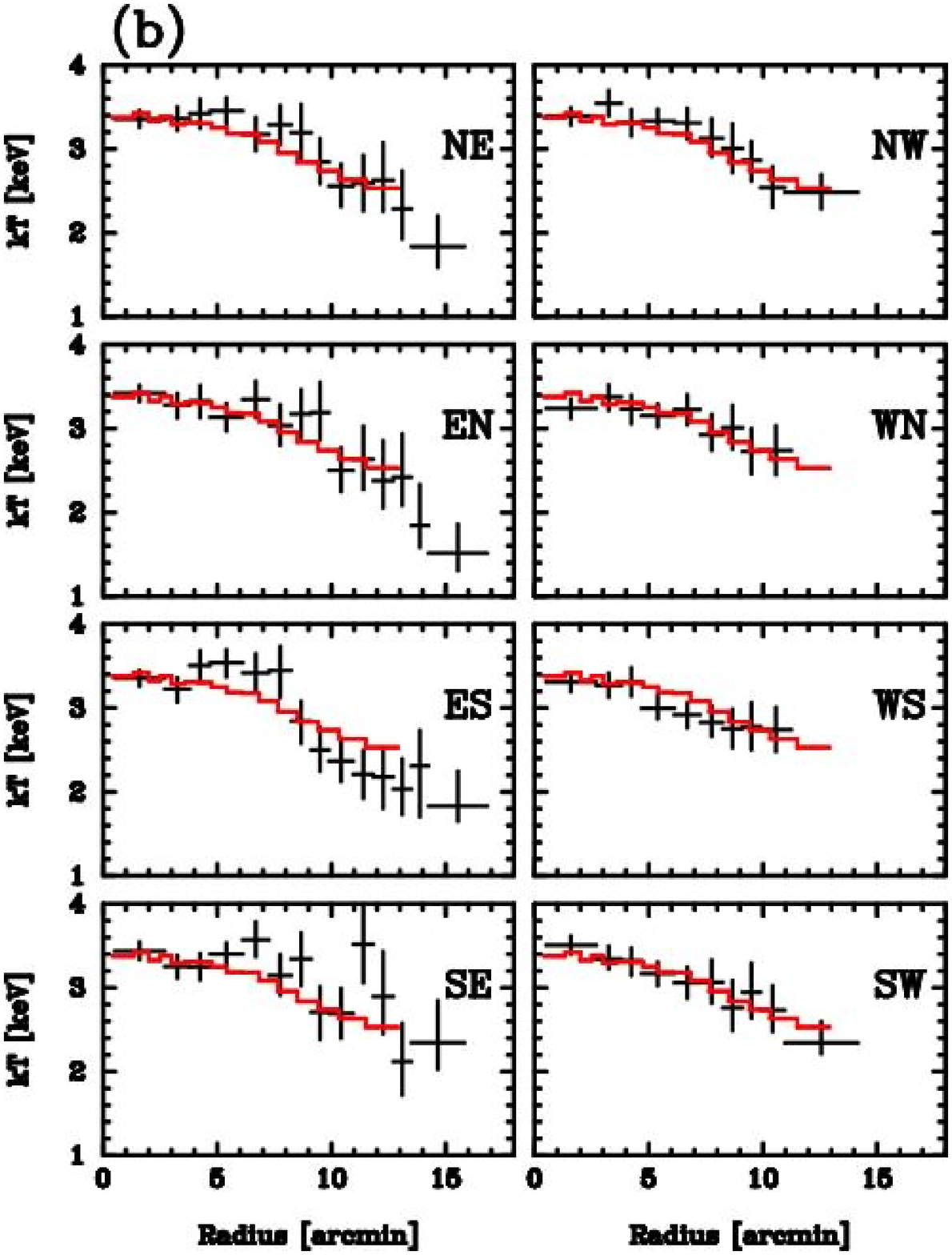}
  \end{center}
  \caption{(a):Temperature map obtained from spectral fits.
The overlaid contours show smoothed X-ray surface brightness. 
The three lines indicate MOS1(red), MOS2(green), and pn(blue) 
fields of view. 
(b):Radial temperature profiles of the 8 sectors shown in (a). 
The red line shows the azimuthally-averaged radial temperature profile
shown in Fig.\ref{fig:radi} (top panel), ignoring errors.}
\label{fig:map-kT}
\end{figure}

By taking full advantage of large effective area of {\it XMM-Newton},
we have attempted to
draw a two-dimensional temperature map on the basis of spectral analysis.
We divided each annulus which was adopted to obtain
the radial profiles in Fig.~\ref{fig:radi}
evenly into 8 azimuthal sectors.
The radial width of annuli was determined by requiring a minimum 
of 8000 counts for all the eight sectors.
The spectra were fitted, after background subtraction,
with a single-temperature MEKAL model with the galactic absorption.
This process is the same as that described in 3.1.
Figure~\ref{fig:map-kT}(a) and (b) show
the resulting two-dimensional temperature distribution,
and its radial profiles in the eight azimuthal directions, respectively.
For reference, the azimuthally-averaged radial temperature profile
(Fig.\ref{fig:radi}) is overlaid in panel (b).
It is remarkable that the $r = 4'\!-\!8'$ region
in the two south-eastern sectors (labelled SE and ES in the figure)
seems to be hotter than the other sectors at the same radii.

%%%%%%%%%%%%%%%%%%%%%%%%%%%%%%%%%%%%%%%%
\subsubsection{Metal distribution}
%%%%%%%%%%%%%%%%%%%%%%%%%%%%%%%%%%%%%%%%

%Figure \ref{fig:map-Z} shows the metal abundance distribution
%contemporaneously obtained with the temperature distribution
%(\S\S~3.2.1).
%%
%\begin{figure}
%  \begin{center}
%    \FigureFile(80mm,80mm){figure/figure6a.eps}
%    \FigureFile(80mm,80mm){figure/figure6b.eps}
%  \end{center}
%\caption{The same as Fig. \ref{fig:map-kT}, but for the metal abundance. }
%\label{fig:map-Z}
%\end{figure}
%%
%Due to statistical limitation, however,
%no definitive feature is found
%in the azimuthally-segmented radial metallicity distribution
%that deviates from the averaged one.

\begin{figure}
  \begin{center}
    \FigureFile(80mm,80mm){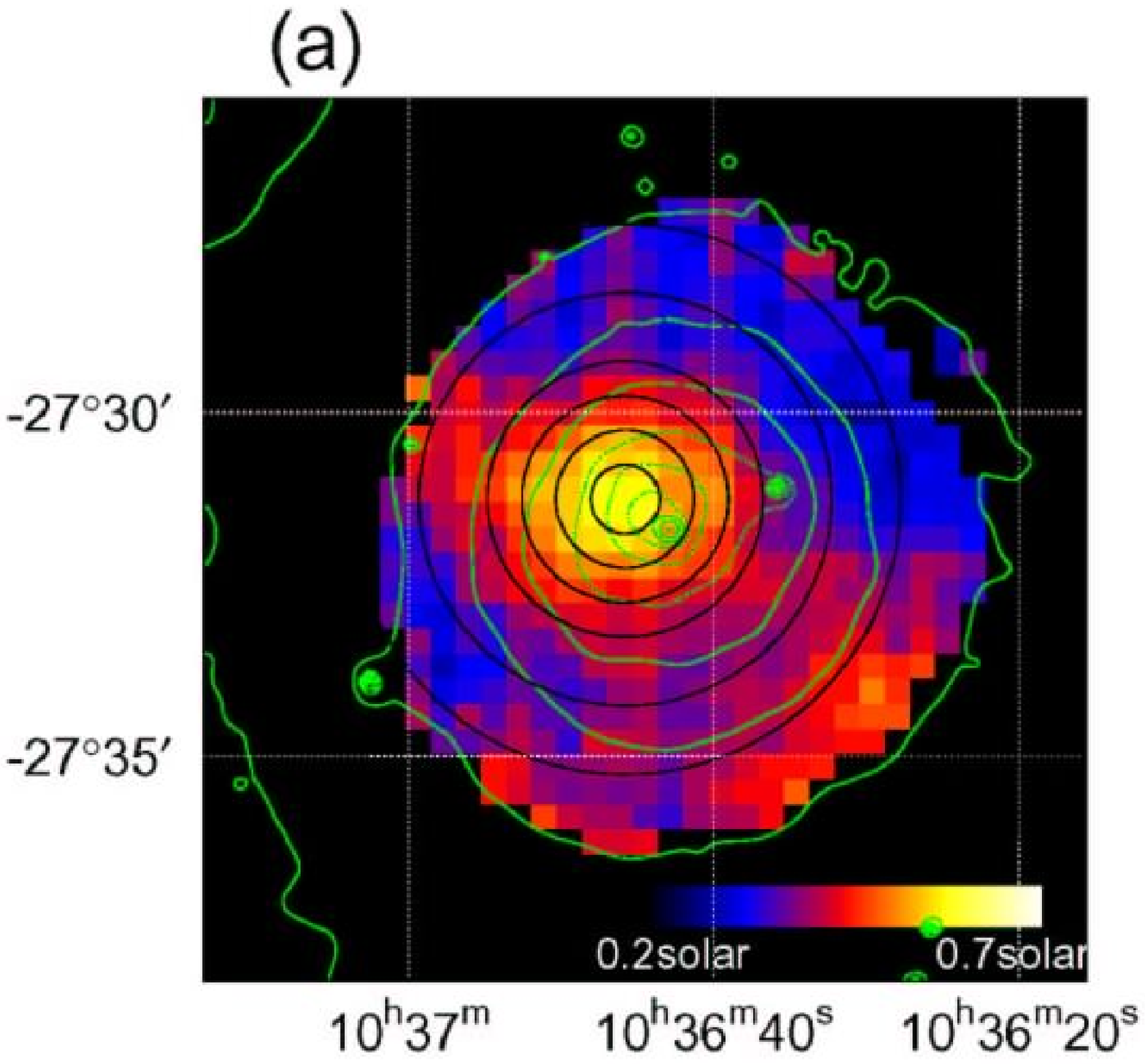}
    \FigureFile(80mm,80mm){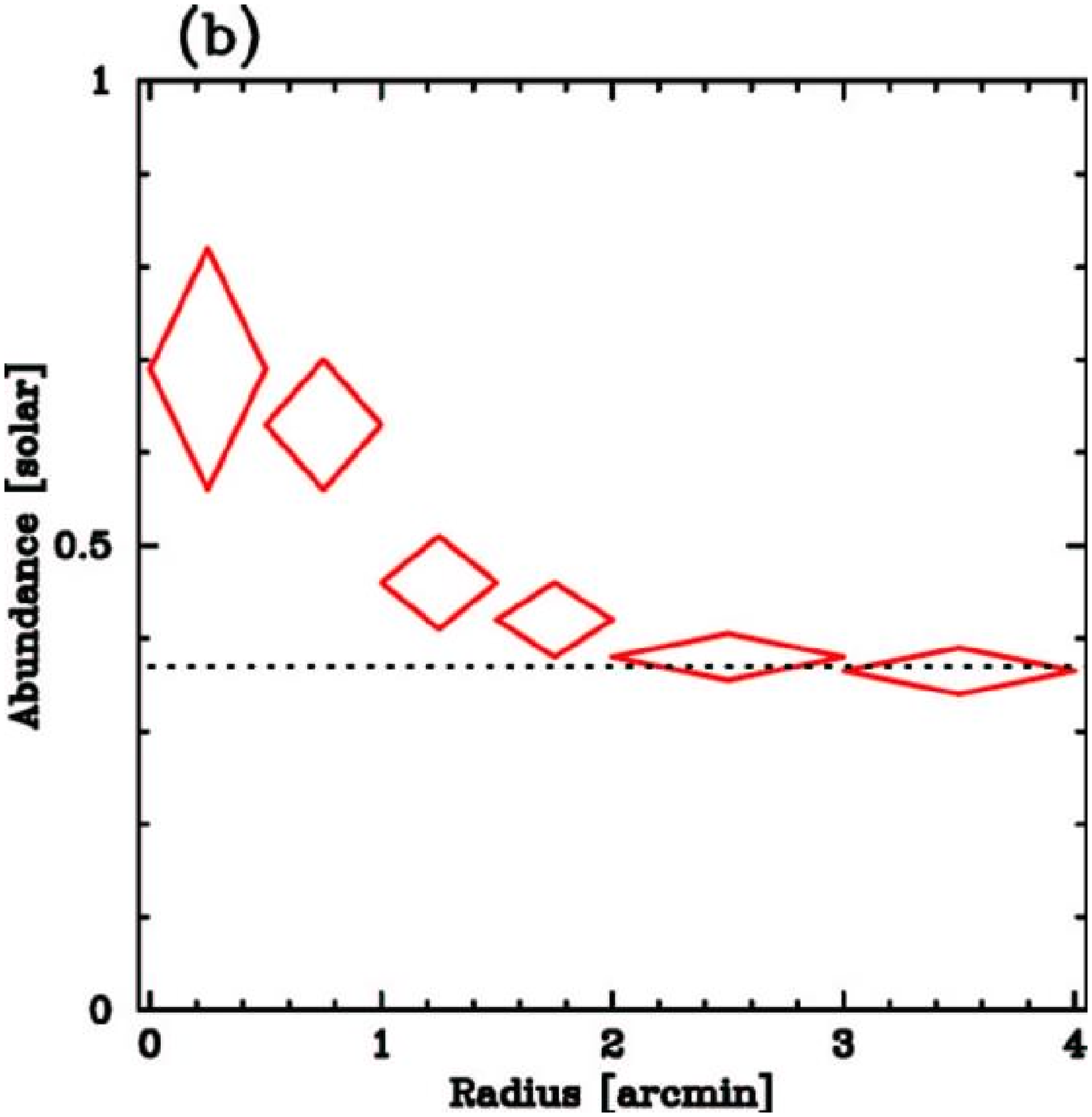}
  \end{center}
\caption{(a) Metal abundance distribution in the central $r<5'$
region based on spectral analysis.
The pixel size is $22''\times 22''$ (5kpc across). 
Spectral analysis is carried out for a square region 
of $3 \times 3$ pixels.
The overlaid contours show smoothed X-ray surface brightness. 
(b) Radial metallicity distribution in annuli 
centered on the position of the high-metallicity blob,
according to the concentric annuli of (a). 
\label{fig:gridZ}
}
\end{figure}

We reported several high-metallicity regions near the center of A~1060
cluster ($r<5'$) based on our {\it Chandra} observation
\citep{2004PASJ...56..743H}.
We thus attempted to produce fully two-dimensional metal abundance map
in the central $r\lesssim 4'$ region.
We divided the sky region into rectangular cells with $22''$ on a side,
and evaluated the metal abundance by fitting MEKAL model
to the spectra obtained from $3 \times 3$ cells.
We have obtained two-dimensional metallicity distribution
through taking running average 
by shifting the integration region stepped by one cell.
In order to keep statistical quality,
we discard any $3 \times 3$ cell
if it does not contain at least 4000 counts.
The result is shown in Figure \ref{fig:gridZ}(a).
A typical statistical error of the metallicity is 10--15\%.
We note that there is clearly a high-metallicity region 
at $\sim\! 1.\!\!'5$ northeast of NGC~3311 whose position is consistent
with the blob detected with the {\it Chandra} observation.
The projected metallicity within a radius of $1'$
centered on the blob is $0.65\pm 0.06Z_\odot$,
which is compatible with the {\it Chandra} value
of $0.74^{+0.29}_{-0.23}Z_\odot$,
and is about twice as large as in other regions.
Figure \ref{fig:gridZ}(b) shows a radial metallicity distribution 
in the concentric annuli centered on the blob.
Assuming that the metallicity of the ICM out of the blob region is uniform
at $\sim 0.4Z_\odot$,
the deprojected central metallicity of the blob
is obtained to be $\sim$1.5$Z_\odot$,
which is consistent with the {\it Chandra} value
\citep{2004PASJ...56..743H}.
Note, however, that the radial extent of the high-metallicity region
is found as large as $r \simeq 2'$ (Fig.~\ref{fig:gridZ}b),
whereas it was only $\sim$40$''$ in the {\it Chandra} observation.
This is because the brightness of the blob is
low and close to the photon limit of the {\it Chandra} ACIS.
None of the other high-metallicity blob-like regions
suggested by the {\it Chandra} observation,
which are all fainter than the central blob,
was significantly detected in the {\it XMM-Newton} observation.

%%%%%%%%%%%%%%%%%%%%%%%%%%%%%%
\section{Discussion}
%%%%%%%%%%%%%%%%%%%%%%%%%%%%%%

The present {\it XMM-Newton} observation of A~1060
provided several pieces of new knowledge
on the surface brightness, temperature and abundance distributions
in unprecedented detail.
They include the significant drop of the temperature 
toward outer radii (\S~3.1),
the existence of the south-eastern high temperature region (\S~3.2.1),
the high-metallicity blob in the central region (\S~3.2.2), and so on.
We discuss their implication in some detail.

%%%%%%%%%%%%%%%%%%%%%%%%%%%%%%
\subsection{Mass Distribution}
%%%%%%%%%%%%%%%%%%%%%%%%%%%%%%

We first derive the gravitational mass profile
based on the density distribution according to the following steps
(see also \S~4.2 of Hayakawa et al. 2004):
\begin{enumerate}
 \item We first calculate the radial surface brightness profile 
   from the MOS1+2 image in the 0.8-8.0~keV band, after application 
   of the SAS task ``evigweight'' in order to eliminate the telescope 
   vignetting effect.
 \item Under the assumption of constant temperature 
   and spherically symmetric distribution of the ICM,
   we fit the radial surface brightness profile in the range $r>5\!'$ 
   with a single $\beta$-model.
   The fitted parameters are $r_{c} = 7.\!\!'3\pm 1.\!\!'8$
   and $\beta = 0.69^{+0.11}_{-0.22}$, respectively.
 \item The best-fit $\beta$-model is extrapolated to the center,
   and the ratio of the data to the model is calculated as a function 
   of the radius.
   The radial profile of the surface brightness ratio is fitted
   with an exponential function.
 \item The density of the plasma $\rho (r)$ predicted by the
       $\beta$-model is enhanced by multiplying 
   the square root of the above ratio, and the new radial profile of 
   the brightness is calculated.
 \item The procedures 3 and 4 are repeated until
   the residual between the data and the model practically vanishes.
\end{enumerate}
As a result, we obtained the central electron number density 
to be $n_{0}=11.7^{+0.7}_{-0.6} \times 10^{-3}$~cm$^{-3}$.
This value is slightly larger than that obtained from
the {\it Chandra} observation
$n_{0}=8.2^{+1.8}_{-1.0} \times 10^{-3}$~cm$^{-3}$
\citep{2004PASJ...56..743H}.
This is mainly because we have substituted the exponential function
for the Gaussian adopted for the {\it Chandra} data
in the procedure 3 above.
Since the statistics of the radial surface brightness profile
of {\it Chandra} data was poor at the cluster center,
the choice of the model to fit to the data-to-model ratio,
especially at the cluster center, is not unique.
We thus simply adopted the Gaussian with no particular reason.
In analysing the high quality {\it XMM-Newton} data, however,
we have found that the exponential function is significantly
better than the Gaussian.

\begin{figure}
  \begin{center}
    \FigureFile(80mm,80mm){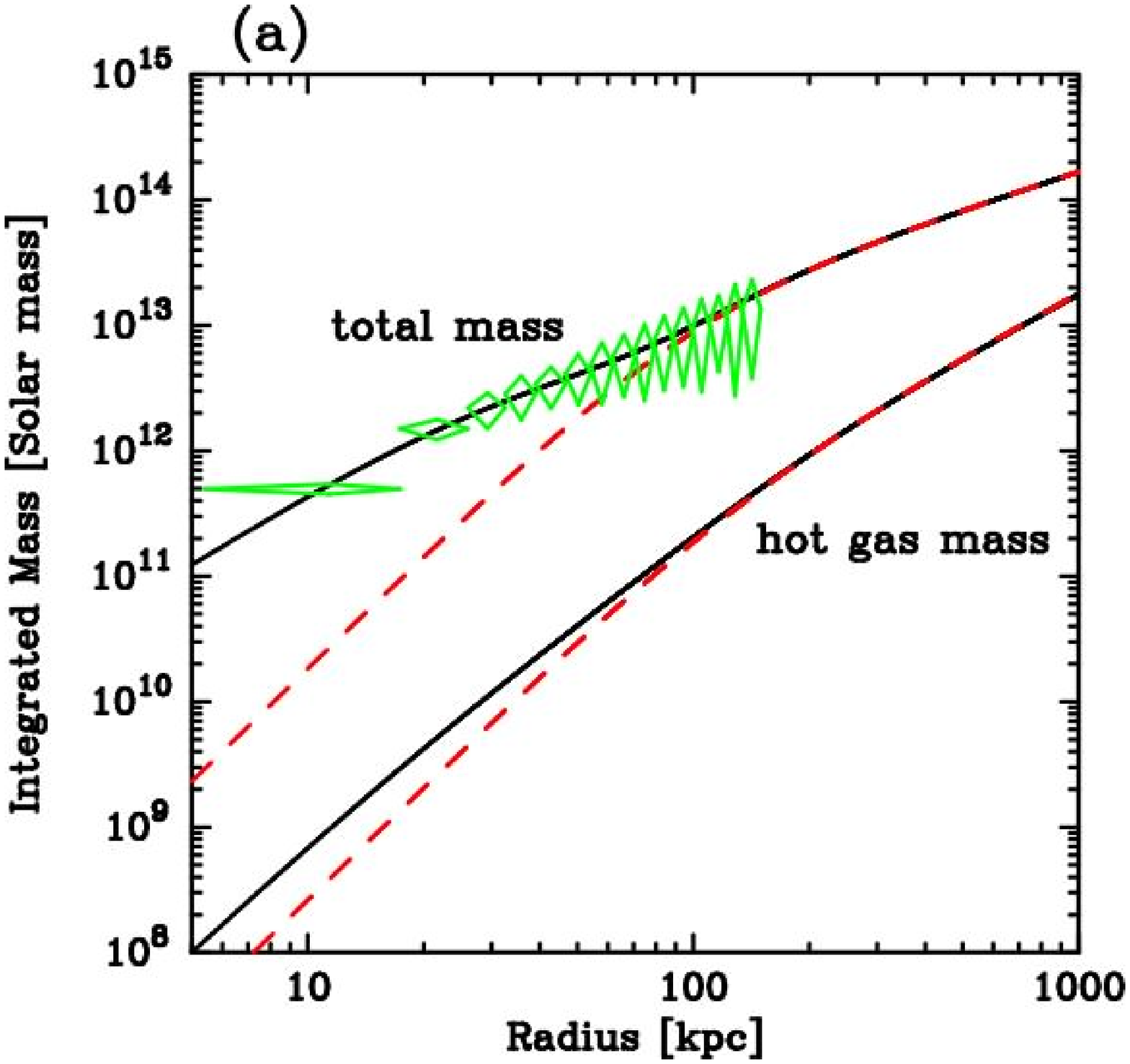}
    \FigureFile(80mm,80mm){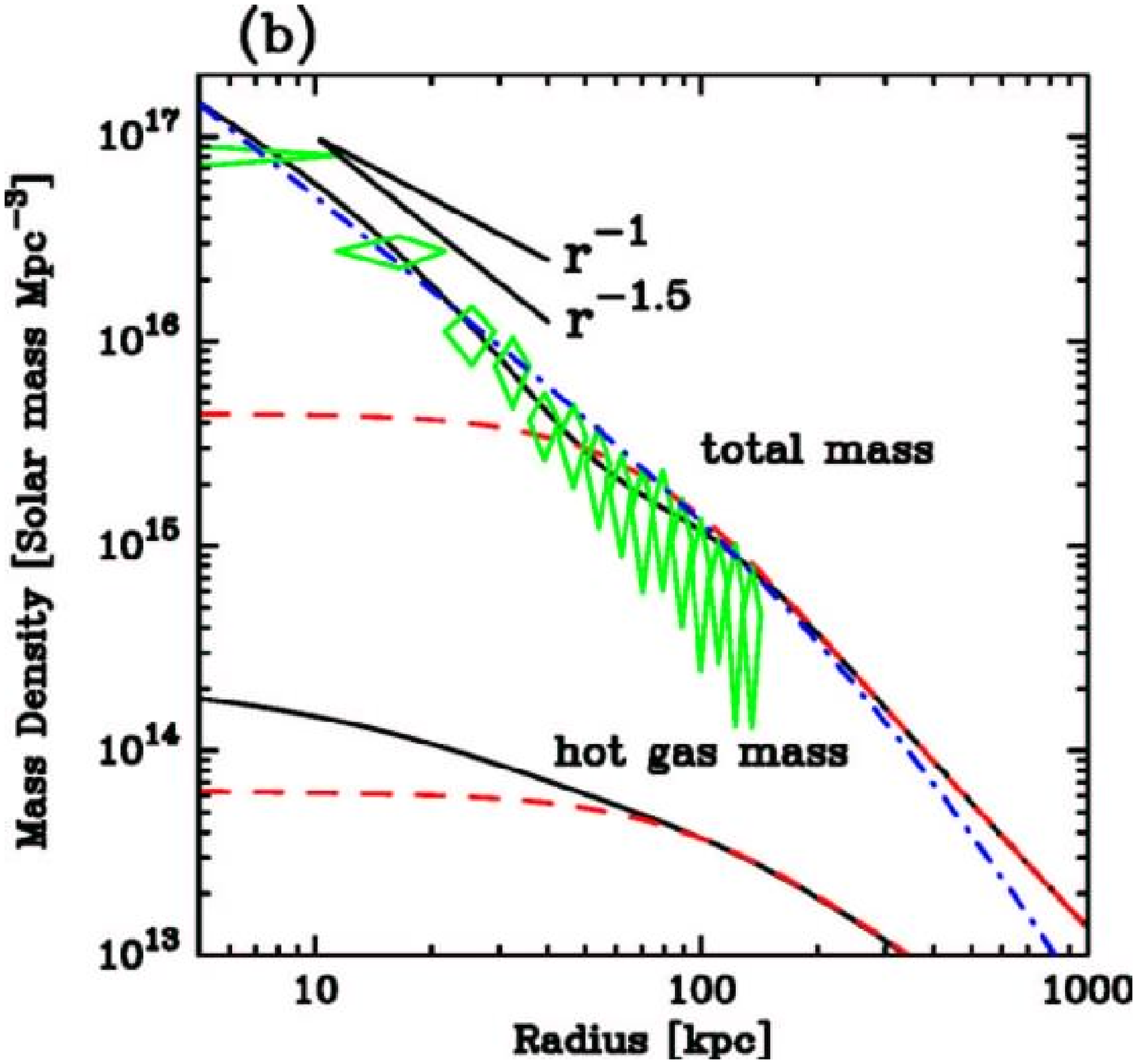}
  \end{center}
\caption{(a) Gravitational mass profile estimated
from the single $\beta$-model (upper dashed line)
and the one obtained from our analysis (upper solid line)
under the assumption of the isothermal ICM in hydrostatic equilibrium.
The lower lines represent ICM masses only.
The radial temperature gradient (Fig.~\ref{fig:radi}) is reflected
for the diamonds.
(b) Gravitational mass density profile based on (a).
The dashed-dotted line shows the Moore's model.
Two representative slopes, $r^{-1.5}$ and $r^{-1}$, are also shown by
 solid lines.
%with a central density slope of 1.5.
\label{fig:mass}}
\end{figure}

Given the radial density profile that can reproduce
the surface brightness distribution, 
we then calculate the total gravitational mass with the equation
\begin{displaymath}
M_{tot}(<r)=-\frac{rkT(r)}{\mu m_{p} G} \left(
   \frac{d \ln \rho(r)}{d \ln r} + \frac{d \ln T(r)}{d \ln r}
\right)
\end{displaymath}
under to assumption of the hydrostatic equilibrium 
and the spherical symmetry of ICM.
The integrated gravitational and gas mass profiles
are shown by the solid lines in Fig.~\ref{fig:mass}(a)
under the assumption of the constant
temperature $T(r) = \langle T \rangle$ (=3.3~keV; \S~2).
The corresponding mass density profiles are also shown
in Fig.~\ref{fig:mass}(b).
For comparison, those
assuming the single $\beta$-model are plotted with the red dashed lines.
Shown by diamonds are the profiles
that reflect the radial temperature variation (Fig.~\ref{fig:radi}).
Owing to large effective area of {\it XMM-Newton},
we have unveiled the significant concentration of mass
in the central region that can be represented
by a density cusp with $\propto r^{-1.5}$.
In order to confirm this,
we tried a brief model fitting to the observed profile
in Fig.~\ref{fig:mass}(b).
We tried two analytical models; 
the NFW model \citep{1997ApJ...490..493N} and 
the Moore's model \citep{1999MNRAS.310.1147M}.
They have a central density cusp
proportional to $r^{-1}$ and $r^{-1.5}$, respectively.
As a result, we found that Moore's profile
could be better fit to the data,
which is displayed in Fig.~\ref{fig:mass}(b) with the dash-dotted line.
The central cusp of the gravitational mass density profile in A~1060
was first suggested by \citet{2000ApJ...535..602T}
based on the analysis of the {\it ASCA} and {\it ROSAT} data.
Their analysis, however, did not take into account the compact halo
associated with the elliptical galaxy NGC~3311
which could raise the cusp-like density enhancement.
In the analysis of {\it Chandra} data,
we could remove contribution from NGC~3311.
Poor statistics, however, precluded us from detecting the density cusp
of the total mass \citep{2004PASJ...56..743H}.

The gravitational mass profile allows us to compute $r_{\Delta}$ 
for an overdensity $\Delta$ using
\begin{displaymath}
\Delta = \frac{3 M_{tot}(<r_{\Delta})}{4 \pi \rho_{c} r_{\Delta}^{3}},
\end{displaymath}
where $\rho_{c}$ is the critical density \citep{2005A&A...433..101P}.
In the SCDM model, the virial radius corresponds to $\Delta=180$.
Our total mass distribution indicates $r_{180}$ to be about 1.35~Mpc.
This is in good agreement with the theoretical prediction
by \citet{1996ApJ...469..494E}, $r_{180} = 2.6$~Mpc~($T_{X}$/10~keV)$^{1/2}$,
giving $r_{180}=1.49$~Mpc for $T_{X}=\langle T\rangle =3.3$~keV,
as the mean emission-weighted temperature.
We hereafter adopt $r_{180}=1.35$~Mpc.

%%%%%%%%%%%%%%%%%%%%%%%%%%%%%%%%%%%%
\subsection{Abundance Distribution}
%%%%%%%%%%%%%%%%%%%%%%%%%%%%%%%%%%%%

The metal abundance distribution in the central $r< 5'$ region
(Fig.~\ref{fig:gridZ}) indicates a high-metallicity region
centered at $\sim\!1.\!\!'5$ northeast of NGC~3311.
This enhancement was identified in our {\it Chandra} paper 
as the metal rich blob with a radius of $40''$ (= 8.7~kpc).
We estimated the total mass of iron contained in the blob to be 
$3.0 \times 10^{6}$~M$_{\odot}$.
{\it XMM-Newton} data showed, however, that the size 
of the high-metallicity region is significantly more extended
than the {\it Chandra} 
result even considering an image blur of the {\it XMM-Newton} data.
This is probably due to the photon limit of {\it Chandra} data 
at the iron line energy band 6--7~keV.
We thus re-estimated the mass of iron contained within a radius of $1.\!'5$ 
($\simeq 20$~kpc) from the blob center.
As we already mentioned in \S~3.2.2, the projected iron abundance
measured with {\it XMM-Newton} is $0.65\pm 0.06Z_\odot$, which is consistent
with the {\it Chandra} value of $0.74^{+0.29}_{-0.23}Z_\odot$.
After the deprojection as we did for the {\it Chandra} data 
\citep{2004PASJ...56..743H}, the total mass of iron in the blob
is estimated to be $1.9 \times 10^{7}$~M$_{\odot}$,
which now becomes larger roughly by an order of magnitude.

The stellar mass of NGC~3311 is estimated 
to be $4.8 \times 10^{11}$~M$_{\odot}$ 
from the relation $M/L_{B}=8.5(M_{\odot}/L_{\odot})$
with $m_B=11.15$ \citep{1989ApJS...69..763F}.
Assuming that the average iron abundance in the stellar mass of NGC~3311
is 1$Z_\odot$, we obtain the total stellar iron mass
as $\sim$1.3 $\times 10^{9}$~M$_{\odot}$.
As a result, the iron mass of the northeast blob is 
$\sim 1.5$\% of the stellar iron content, and
can reasonably be supplied by the single galaxy NGC~3311.
\citet{2005A&A...435L..25S} simulated metallicity evolution in galaxy 
clusters which includes interaction between the galaxies and the ICM.
They have noted that the ram-pressure stripping is more efficient
than the galactic wind in transferring 
metal-enriched interstellar medium into the ICM, 
and the expelled metals are not mixed immediately with the ICM.
The northeast high-metallicity region,
likely to be associated with NGC~3311,
is reminiscent of the ram-pressure stripping from the X-ray image.
If this is the case, NGC~3311 is moving with a velocity larger than a
few 100 km$^{-1}$~s in the central region of A~1060.
The ram-pressure stripping is also likely to be the reason
why this central galaxy shows a very compact X-ray halo
\citep{2002ApJ...578..833Y}.

%%%%%%%%%%%%%%%%%%%%%%%%%%%%%%%%%%%%%
\subsection{Temperature Distribution}
%%%%%%%%%%%%%%%%%%%%%%%%%%%%%%%%%%%%%

Plasma temperature of the A~1060 cluster decreases from the center
to $r = 13'$ by $\sim$30~\%, which is a larger decrement than that detected 
in our {\it Chandra} observation ($\sim$20~\%).
This is probably because background uncertainty
of the {\it Chandra} observation in the outer $r\gtsim 10'$ region.
Our result is consistent with that of
\citet{2000ApJ...535..602T} claiming that the temperature
decreases down to about 2.4~keV at a radius of $\sim 12'\!-\!20'$
based on their analysis of the {\it ASCA} and {\it ROSAT} data.

\begin{figure}
  \begin{center}
    \FigureFile(80mm,80mm){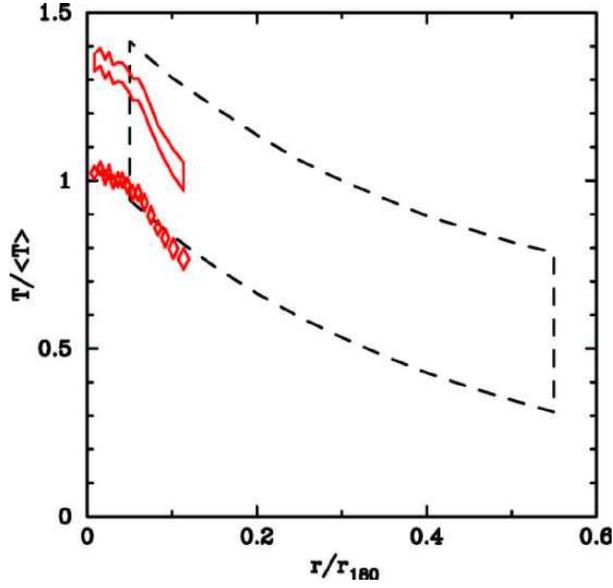}
  \end{center}
\caption{Temperature distribution as a function of the radius 
from the cluster center in a unit of $r_{180}$.
%normalized by the emission-weighted average temperature $<T>$ of 3.3~keV. 
%The red and blue diamonds correspond to those in Fig.~\ref{fig:radi}.
The diamonds and solid line correspond to the red diamonds in
 Fig.~\ref{fig:radi}, normalized by the emission-weighted average
 temperature $<T>$ of 3.3~keV and 2.5~keV, respectively. 
The dashed line define a region that encloses most temperature profile
 and their error bars of \citet{1998ApJ...503...77M}.}
\label{fig:r180}
\end{figure}

\citet{1998ApJ...503...77M} observed 30 nearby clusters
and obtained remarkably similar trend of the temperature
normalized by the emission weighted average temperature $\langle T \rangle$,
as a function of the radius normalized by $r_{180}$.
We thus have attempted to plot our temperature profile
with $r_{180} = 1.35$~Mpc (\S~4.1) 
and $\langle T\rangle = 3.3$~keV (\S~2).
The resulting temperature profile is shown by diamonds 
in Fig.~\ref{fig:r180},
while the composite profile of \citet{1998ApJ...503...77M} is shown 
as the region surrounded by the dashed line.
The profile of A~1060 obviously becomes out of the allowed range
in the region $r/r_{180} \gtsim 0.07$.
One possible reason to cause such a deviation may be an extra heating 
in the central region, due to a sub-cluster merger, for example.
If we are allowed to take $\langle T \rangle = 2.5$~keV,
for example,
the normalized temperature profile well fits in
the Markevitch's range, as displayed in Fig.~\ref{fig:r180}.

\begin{figure}
  \begin{center}
    \FigureFile(80mm,80mm){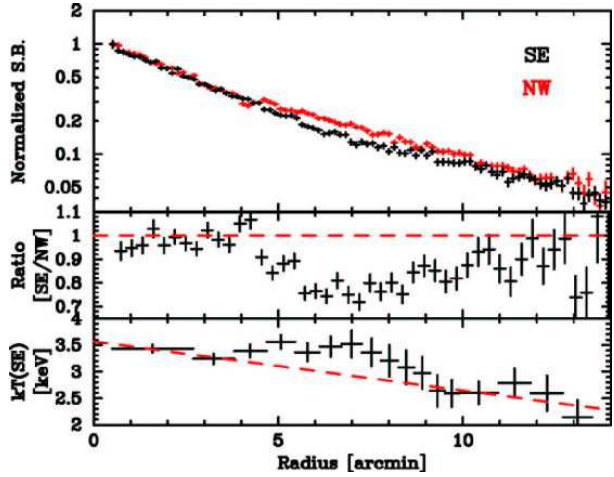}
  \end{center}
\caption{Radial surface brightness profile in the SouthEast direction
in comparison with the NorthWest (NW:red) (top), 
the ratio of these two profiles (middle), 
and the temperature distribution in SE direction (bottom).
The dashed line in the bottom panel is the linear fit to the data
in the range $r<3'$ and $r>10'$.}
\label{fig:quad-radi}
\end{figure}

We also recognize that there is a high temperature region at $r \simeq 7'$ 
from NGC~3311 in the southeast direction (Fig.~\ref{fig:map-kT}).
This feature can be confirmed also by the color-coded temperature map
based on the HR analysis in \citet{2001PASJ...53..421F}.
Figure \ref{fig:quad-radi} shows the radial surface brightness profile
of the southeast direction, compared 
with that in the northwest direction in the top panel. 
The ratio of the two profiles (middle panel),
and the radial temperature profile in the southeast direction 
(bottom panel) are also shown.
The surface brightness around the position of the southeast hotter
region is slightly fainter than that in the northwest region.
The pressure $P$ and the emissivity $I$ are represented
by the electron density $n_{\rm e}$ and the temperature $T$
as $P \propto n_{e} T$ and $I \propto n_{e}^{2} T^{1/2}$. 
Combining these relations by eliminating $n_{\rm e}$, 
we obtain $P \propto I^{1/2} T^{3/4}$.
At the southeast high temperature region, 
the surface brightness is lower than the northwest by $\sim 25$\%,
while the temperature exceeds the linearly-fitted value by $\sim 20$\%.
Thus, the southeast region is in pressure equilibrium
with the environment.
The diffusion time of pressure gradient is 
of the same order as the sound-crossing time.
The time for reaching pressure equilibrium over the scale $3'$ ($\simeq
70$~kpc) can be estimated to be about $4 \times 10^{7}$~yr, for the
sound speed of $\sim 1000$~km~s$^{-1}$.
On the other hand, the temperature difference of 1~keV
over $3'$ distance is relaxed by thermal conduction
whose time scale is estimated to be about $3 \times 10^{8}$~yr
in the case of A~1060.
It is likely that the temperature structure has been produced
between $3 \times 10^{8}$~yr and $4 \times 10^{7}$~yr in the past.

%%%%%%%%%%%%%%%%%%%%
\section{Conclusion}
%%%%%%%%%%%%%%%%%%%%

We have presented the {\it XMM-Newton} observation
of the A~1060 cluster of galaxies.
Large effective area of {\it XMM-Newton} enables us
to investigate spatial distribution of the surface brightness, 
the temperature and the metal abundance in unprecedented detail.
We have derived a gravitational mass profile 
based on a model-independent estimation of the density profile.
The overall shape can be well represented by Moore's profile
with a central density slope $\propto r^{-1.5}$.
The temperature is found to decrease undoubtedly by 30~\% 
from the center to the edge of the field of view $r = 13'$.
This drop is clearly steeper than the Markevitch's composite profile.

Our spatially-resolved analysis revealed slight temperature enhancement
at $r \simeq 7'$ from NGC~3311 in the southeast direction.
It is thought that some heating, like a subcluster merger, 
has occurred in this region in the past 
between the sound crossing time $4 \times 10^{7}$~yr 
and the thermal conduction time $3 \times 10^{8}$~yr ago.
We note that the deviation of temperature profile of A~1060
from the Markevitch's one can also be explained 
if the central region is heated
like the way seen in the southeast region.

The high resolution metallicity map for the central $r< 5'$ region
confirmed the high-metallicity blob
discovered with the {\it Chandra} observation
locating at around $1.\!\!'5$ northeast of NGC~3311.
Its spatial extent is, however, found to be as large as $r \sim 2'$,
which is significantly larger than that estimated with {\it Chandra}
($r\sim 40''$).
This discrepancy is probably because the {\it Chandra} observation
is photon-limited in the iron line band.
The iron mass in this region is revised to be
$1.9 \times 10^{7}$~M$_{\odot}$.
Although this is larger than the {\it Chandra}
value roughly by an order of magnitude, 
it can still be supplied solely by NGC~3311.

%%%%%%%%%%%%%%%%%%%%%%%%%%%%%%%%%%%%%%%

%%%\appendix

\end{document}